\documentclass[11pt]{article}

\let\ssection=\section
\renewcommand{\section}{\setcounter{equation}{0}\ssection}

\newcommand{\half}{{\scriptstyle{\frac{1}{2}}}}

\newcommand{\IR}{{\bf R}}

\newcommand{\vx}{{\vec x}}
\newcommand{\red}{{\rm red}}
\newcommand{\const}{\mathop{\rm const}\nolimits}
\def\parag{\hfil\break} 
\def\kikezd{\parag\underbar}
\def\p{{\partial}}

\def\vgamma{{\vec\gamma}}
\def\vE{{\vec E}}

\def\vp{{\vec p}}

\def\vx{{\vec x}}

\def\vE{{\vec E}}

\begin{document}

\setlength{\baselineskip}{16pt}

\title{The non-commutative Landau problem\\[8pt]
}
\author{
P.~A.~Horv\'athy
\\
Laboratoire de Math\'ematiques et de Physique Th\'eorique\\
Universit\'e de Tours\\
Parc de Grandmont\\
F-37 200 TOURS (France)
}

\date{\today}

\maketitle

\begin{abstract}
     The Landau problem is discussed in two similar but still different
     non-commuta\-tive frameworks. The ``standard'' one, where the
     coupling to the
     gauge field is achieved using Poisson brackets, yields
     all Landau levels. The  ``exotic'' approach, where the coupling to the
     gauge field is achieved using the symplectic structure,
     only yields lowest-Landau level states,
     as advocated by Peierls and used in
     the description of the ground states of the Fractional Quantum Hall
Effect.
     The same reduced model also describes
     vortex dynamics in a superfluid ${}^4$He film.
     Remarkably, the spectrum  depends crucially on the quantization
     scheme. The system is symmetric w. r. t. area-preserving diffeomorphisms.
\end{abstract}

\noindent
{\sl Ann. Phys.} (N.Y.) {\bf 299}, 128-140, 10 July 2002

\goodbreak

\section{Introduction}

The non-commutative Landau problem,
(i.e., a particle in the non-commutative plane,
coupled to a constant magnetic field $B$ and an electric potential
$V$) has genereted a  considerable amount of recent interest
\cite{DJT, DuJa, DH, NaPo, GAMB, Sochi, BNS}. The starting point of
the ``standard'' approach \cite{NaPo, GAMB, Sochi, BNS} is
to consider the commutation relations and Hamiltonian
\begin{equation}
     \begin{array}{ll}
	&[x_{1}, x_{2}]=i\theta,
	\qquad
	[x_{i},p_{j}]=i\delta_{ij},
	\qquad
	[p_{1}, p_{2}]=iB,\hfill
	\\[10pt]
	&H=\half{\vp^2}+V\hfill
  \end{array}
  \label{BNSrel}
\end{equation}
and observe that the
behaviour of the system depends qualitatively on the sign of the
parameter
\begin{equation}
     m^*=
     1-B\theta.
\label{BNSeffmass}
\end{equation}
When $m^*=0$, the representation of the Heisenberg algebra of the
$x_{i}$ alone becomes irreducible \cite{Sochi}, so that the Casimirs
are constants (chosen to be zero in \cite{BNS}),
\begin{equation}
     \pi_{i}=p_{i}-\frac{\varepsilon_{ij}x_{j}}{\theta}=0.
     \label{casimir}
\end{equation}
  Then the problem becomes explicitly solvable;
the exact Landau-type energy spectrum is
$
E_{n}=\theta^{-1}(n+\half)+\epsilon_{n},
\;  n=0,1,\dots,
$
where $\epsilon_{n}$ is the eigenvalue of the potential
alone \cite{BNS} (further discussed below).
\goodbreak


Non-commuting coordinates have arised before
in the study of the ground states of the Fractional Quantum
Hall Effect \cite{QHE, GJ}~: seventy years ago,
Peierls \cite{Peierls} argued in fact that
   in a strong  magnetic field and a sufficiently
  weak  potential the lowest  Landau level retains its identity.
Putting the mass and the electric charge  again
  to unity, the energy eigenvalues are
$
E_{n}={B}/{2}+\epsilon_{n},
$
where the
$\epsilon_{n}$ are the eigenvalues of the operator
$\hat{V}(\hat{p},\hat{q})$, obtained from the
potential alone, but such that $\hat{p}$ and $\hat{q}$ are
canonically conjugate,
$
[\hat{p},\hat{q}]=-i/B.
$

This ``Peierls substitution'' can be justified taking the
$m\to0$ limit of ordinary quantum mechanics \cite{DJT, DuJa}.
A slightly different derivation
  \cite{DH}  starts with  a classical particle
  associated with the ``exotic'' two-parameter  central
extension of the planar Galilei group \cite{exotic}.
The extension parameters combine with the magnetic field
into and effective mass, which is, indeed,
  the parameter $m^*$  in (\ref{BNSeffmass});
when this latter
vanishes, the Peierls substitution is once again recovered
using Hamiltonian reduction \cite{FaJa}.

Despite their deceptive similarity, the standard
  NC  \cite{NaPo, GAMB, BNS} and the Peierls-type ``exotic''
  \cite{DH} approaches are, nevertheless, {\it different}~:
in the critical case $m^*=0$,
the spectrum of \cite{BNS}
   exhibits all Landau levels, while the Peierls spectrum \cite{DJT,
   DuJa, DH}
  only consists of LLL (Lowest Landau Level) states.

Analyzing the strong similarities and subtle differences of
the two approaches using a unified framework, we find that the
``standard'' coupling is not entirely satisfactory,
while the ``exotic'' approach works perfectly.

A major issue is the spectrum~:
surprisingly, it depends crucially on the quantization scheme
we use, and we have not been able to find the ``good'' answer.

A further remarkable fact is that the classical counterpart of the
Peierls model \cite{DJT, DH} also
describes another, related, but physically different, situation, namely
the effective dynamics of vortices in a thin superfluid
${}^4$He film \cite{HMC, LeMy}. Our remarks provide us with yet
another indication for the intimate
relationship of supefluidit and the Hall effect.
Physically, the ``exotic'' model describes
Laughlin's quasiparticles,  which in turn correspond
to the vortices of the field theory \cite{QHE}.

\section{The spectra}

Firstly, we rederive the spectra in a unified framework.
Let us first assume that $V$ (but not $B$) vanishes,
$V=0$, $B\neq0$.

\subsection{The ``standard'' NC model}

Let us start with the standard NC approach \cite{NaPo, GAMB, BNS}.
Let  $m^*\neq0$.
The classical counterparts
of the commutation relations obtained by
replacing (\ref{BNSrel}) by Poisson brackets
are associated with the symplectic form
\begin{equation}
     \Omega_{NC}=\frac{1}{m^*}\Big[
     d\vp\wedge d\vx+\theta dp_{1}\wedge dp_{2} +
     B dx_{1}\wedge dx_{2}\Big].
     \label{NCsymp}
\end{equation}
Note that the condition $d\Omega_{NC}=0$
requires the magnetic field $B$ to be constant.
Introducing the complex coordinates
\begin{equation}
     \begin{array}{ll}
     z=\displaystyle\frac{1}{\sqrt{m^*}}\left(
     \sqrt{B}\big(x_{1}+ix_{2}\big)
     +
     \frac{1}{\sqrt{B}}\big(-ip_{1}+p_{2}\big)\right),
     \hfill\\[8pt]
     w=\displaystyle\frac{1}{\sqrt{B}}\big(-ip_{1}-p_{2}\big),\hfill\\
     \end{array}
     \label{NCcomplcoord}
\end{equation}
(\ref{NCsymp}) is rewritten as
$
  \Omega_{NC}=(2i)^{-1}\big({d\bar{z}\wedge dz+ d\bar{w}\wedge dw}\big),
$
which shows that $z$ and $w$ are canonical coordinates
on phase space.
Note that these definitions ``mostly use'' the magnetic field
$B$; the non-commutative paremeter $\theta$ only enters $z$, namely
through the  pre-factor $(m^*)^{-1/2}$.
Then, choosing the  antiholomorphic polarization
yields the Bargmann-Fock wave functions
$f(z,w)e^{-(\vert z\vert^2+\vert w\vert^2)/4}$
with $f(z,w)$ holomorphic in both variables.
Expressed as acting on the holomorphic functions alone,
the fundamental operators read
\begin{equation}
    \begin{array}{ll}
    \widehat{z}=z\cdot
    \qquad
    &\widehat{\bar{z}}=2\p_{z}\hfill
    \\[4pt]
    \widehat{w}=w\cdot
    \qquad
    &\widehat{\bar{w}}=2\p_{w}\hfill
    \end{array}
    \label{zwop}
\end{equation}
\noindent
and satisfy
\begin{equation}
	[\widehat{\bar{z}},z]=2=[\widehat{\bar{w}},\widehat{w}],
	\qquad
	[\widehat{z},\widehat{w}]=0=[\widehat{\bar{z}},\widehat{\bar{w}}].
\label{zwcommrel}
\end{equation}
The classical kinetic Hamiltonian,
$H_{0}=\vp^2/2=\frac{B}{2}w\bar{w}$,  is quantized into
\begin{equation}
     \widehat{H}_{0}=\half B\big(\widehat{w}\widehat{\bar{w}}+1\big)
     =B\left(w\p_{w}+\half\right).
     \label{NCkinham}
\end{equation}

Now if we let $m^*$ go to zero, both the symplectic form
$\Omega_{NC}$ and
the coordinate $z$ blow up. Regularity can be maintained, though,
if we require that
\begin{equation}
     z=0.
     \label{NCcondition}
\end{equation}
Therefore, $f$ is a function of $w$ alone.
(Note that $w$ is essentially $\vp$).
Remarkably, $z=0$ is precisely the condition of the vanishing
of the Casimirs, $\pi_{i}=0$,  in (\ref{casimir}) \cite{BNS, Sochi},
since $B=\theta^{-1}$ in the critical case.
The eigenfunctions,  $w^n$, have
eigenvalues $E_{n}=B(n+1/2$), i. e.,  the first term in
the spectrum in \cite{BNS}.

\subsection{The exotic model}

Let us now turn briefly to the ``exotic'' model;
for more details the Reader is referred to \cite{DH}.
The symplectic form is simply
\begin{equation}
     \Omega_{E}=
     d\vp\wedge d\vx+\theta dp_{1}\wedge dp_{2} +
     B dx_{1}\wedge dx_{2},
     \label{Esymp}
\end{equation}
but the associated Poisson brackets
acquire an $(m^*)^{-1}$ factor:
\begin{equation}
	\{x_{1}, x_{2}\}=\frac{\theta}{m^*},
	\qquad
	\{x_{i},p_{j}\}=\frac{\delta_{ij}}{m^*},
	\qquad
	\{p_{1}, p_{2}\}=\frac{B}{m^*}.
  \label{Erel}
\end{equation}
We modify therefore (\ref{NCcomplcoord}) as
\begin{equation}
     \begin{array}{ll}
     z=\sqrt{B}\big(x_{1}+ix_{2}\big)
     +
     \displaystyle\frac{1}{\sqrt{B}}\big(-ip_{1}+p_{2}\big)
     \hfill\\[10pt]
     w=\displaystyle\sqrt{\frac{m^*}{B\ }}\big(-ip_{1}-p_{2}\big),\hfill\\
     \end{array}
     \label{Ecomplcoord}
\end{equation}
so that  now
$(2i)^{-1}\big({d\bar{z}\wedge dz+ d\bar{w}\wedge dw}\big)=
     \Omega_{E}$.
The new $z$ and $w$ are hence again canonical coordinates, quantized,
for $m^*\neq0$, as in (\ref{zwop}); the relations
(\ref{zwcommrel})  still hold. The kinetic hamiltonian,
$H_{0}=\frac{B}{2m^*}w\bar{w}$, is quantized to
\begin{equation}
     \widehat{H}_{0}=\frac{B}{2m^*}\big(\widehat{w}\widehat{\bar{w}}+1\big)
     =
     \frac{B}{m^*}\big(w\p_{w}+\half\big),
     \label{kinham}
\end{equation}
which differs from (\ref{NCkinham}) in the
pre-factor $(m^*)^{-1}$.

Unlike $\Omega_{NC}$, the symplectic form (\ref{Esymp}) does not
diverge as $m^*\to0$; it becomes, however,
singular,  since det$(\Omega_{E})=(m^*)^2$.
Equivalently, the associated Poisson brackets (\ref{Erel}) diverge.
These divergences can be avoided by eliminating $\bar{w}$,
\begin{equation}
\half\widehat{\bar{w}}f\equiv\p_{w}f=0,
\label{Econd}
\end{equation}
so that $f$ becomes a function of
the ``position'' coordinate $z$ alone: we recover  hence the
Laughlin prescription used in the FQHE context
\cite{QHE}.
Restricting ourselves to the subspace $\p_{w}f=0$
  we find, for $m^*\neq0$,
that  any $f(z)$ has the ground state energy,
\begin{equation}
     E_{0}=\frac{B}{2m^*}.
\label{Egroundenergy}
\end{equation}
All Laughlin wavefunctions belong hence to the LLL.
Their energy diverges as $m^*\to0$, though.
\goodbreak

\subsection{Quantization of the potential}
\label{quantization}

Restoring the potential, we observe that
  the  total Hamiltonian is, in both cases,
\begin{equation}
H\equiv H_{NC}=H_{E}=\half{\vp^2}+V.
\label{hamiltonian}
\end{equation}

  Our task is to quantize it on the respective Hilbert spaces.
  Owing to the non-commutativity of the coordinates,
quantizing the potential $V=V(\bar{z}, z, \bar{w},w)$
is, in general, a rather difficult
problem. We restrict ourselves therefore to the critical case
$B\theta=1$ and assume that the potential is radial,
$V=U(r^2)$.

In the NC model we eliminate the coordinate $z$. Then
$\vert\vx\vert^2=\vert\vp\vert^2/B^2=\bar{w}w/B$, so that
our task is to quantize
$U(\bar{w}w/B)$ as acting on the analytic functions
$f(w)$, subject to $[\widehat{\bar{w}},\widehat{w}]=2$.

In the exotic case instead, it is $w$ which is eliminated and we
are left with analytic functions $f(z)$,
subject to $[\widehat{\bar{z}},\widehat{z}]=2$. Now $V=V(\bar{z},z)
  = U(\bar{z}z)$,
  so the task is,
in both cases, consistent with the kinetic part in the Hamiltonian.
We can treat hence both problems simultaneously,
with $Z$ denoting either of the remaining complex variable,
$w$ or $z$, respectively.
For simplicity, we choose  $B=\theta=1$.

A remarkable fact is now that the quantum operator $\widehat{V}$
depends crucially on the chosen quantization scheme.

Some people \cite{BNS, HMC, GAMB2} claim that
the spectrum is simply $U(\theta(2n+1))$.
This statement can be justified, e. g.,
by excluding the origin and
considering the real polarization $r=\const$. in the plane.
The flow of any function of $r$ preserves this polarization, so that
  $\widehat{V}$ is simply multiplication by $U(r^2)$, and the
spectrum formula follows at once.

Other quantization schemes yield other results, though.
The main difficulty in constructing the quantum operator
$\widehat{V}$ is in fact to resolve
the ambiguity in ordering the non-commutative fundamental
variables. In fact, for an arbitrary
classical observable $A$ with  associated
operator $\widehat{A}$, and an arbitrary real function $U$,
$\widehat{U(A)}\neq U(\widehat{A})$ in general.

Now, according to one proposal \cite{DJT, DuJa, GJ},
$\widehat{V}$ is obtained, in the holomorphic Bargmann-Fock representation,
by {\it anti-normal ordering}, which amounts to ``putting all the
$\widehat{\bar{Z}}$ to
the left and all the $\widehat{Z}$ to the right'' \cite{GJ, DuJa}.
In the radially-symmetric polynomial case
\begin{equation}
     V_{N}=\left(\half{\bar{Z}Z}\right)^N,
    \label{Npot}
\end{equation}
this rule yields the recurrence relation
$
\widehat{V}^{an}_{N}=\widehat{V}^{an}_{N-1}(Z\p_{Z}+N).
$
Hence, using  $\widehat{V}^{an}_{0}$=1,
\begin{equation}
     \widehat{V}^{an}_{N}=(Z\p_{Z}+1)\dots(Z\p_{Z}+N)
     \label{Van}
\end{equation}
The eigenfunction of $\widehat{V}=Z\partial_{Z}+1$
  are $f_{n}=Z^n$ with eigenvalues $n+1$.
  The spectrum of $\widehat{V}^{an}_{N}$ is therefore
\begin{equation}
     \epsilon_{n}^{an}=(n+1)\dots(n+N).
     \label{Jopspect}
\end{equation}

It is easy to see that
the prescription requires modification, though: in the simplest case
$N=1$ i. e. for a quadratic potential,
$V=\half Z\bar{Z}$, for example, $\epsilon_{n}^{an}=n+1$.
  Observing, however, that this $V$ is actually the
  {\it full} Hamiltonian of
  a $1$--{\it dimensional} oscillator with {\it phase-space} variable $Z$,
we conclude that the spectrum should be rather $n+1/2$.
This discrepancy has been noticed by the authors of \cite{DJT}, who,
as they say, have no {\it a priori} determination for it.

Yet another quantization scheme
due to Bergman  \cite{Bergman} identifies
the quantum operator associated to $V(Z,\bar{Z})$ as
\begin{equation}
     \widehat{V}\psi(Z)=
     \int\exp\left[\half\bar{z}(Z-z)\right]
     \left(V-\p_{z}\p_{\bar{z}}V\right)\psi(z)dzd\bar{z}.
     \label{Bergmankernel}
\end{equation}
This rule has also been derived rigorously, using geometric
quantization by Tuynman in Ref. \cite{Bergman}.

When $V$ is a polynomial, the formula simplifies
as follows:  \underline{first} calculate
$
\widetilde{V}=V-\p_{Z}\p_{\bar{Z}}V,
$
  and \underline{then} quantize $\widetilde{V}$ by anti-normal ordering.
In the quadratic case above the correction term
subtracts $1/2$, yielding the correct formula, namely
\begin{equation}
     \widehat{V}=Z\p_{Z}+\frac{1}{2},
         \label{goodosc}
\end{equation}
whose spectrum is $\epsilon_{n}=n+1/2$.
(In this case we agree with \cite{GAMB, BNS}).

More generally, in the radially-symmetric polynomial case with $N\geq2$
we have
\begin{eqnarray*}
\widetilde{V}_{N}=\left(\half{\bar{Z}Z}\right)^N
-\frac{N^2}{2}\left(\half{\bar{Z}Z}\right)^{N-1},
\end{eqnarray*}
from which one infers that
$
\widehat{V}_{N}=\widehat{V}^{an}_{N}
-\frac{N^2}{2} \widehat{V}^{an}_{N}.
$
Bergman quantization yields hence
\begin{equation}
     \widehat{V}_{N}=
     (Z\p_{Z}+1)\dots(Z\p_{Z}+N-1)(Z\p_{Z}+N-\frac{N^2}{2}),
     \label{BergVop}
\end{equation}
whose spectrum is
\begin{equation}
     \epsilon_{n}^{\rm crit}=
     (n+1)\dots(n+N-1)(n+N-\frac{N^2}{2})
     \label{Bergspectrum}
\end{equation}
that disagrees with both previous results,
$U(2n+1)=(n+\half)^N$,
of \cite{BNS}, as well as  (\ref{Jopspect}).

The result is unsatisfactory though~: for $N\geq3$, the
first few eigenvalues in
(\ref{Bergspectrum}) become {\it negative}. This is
absurd, and one would need a rule to exlude those eigenvalues.
Unfortunately, we have not been able to derive such a rule
from first principles. We can not adopt, therefore, any
final answer to the problem of finding the ``good'' quantization.

Returning to our models, the total Hamiltonian
  ${\vp^2}/2+V$ acts, for $m^*\neq0$,
on the ``unreduced'' Hilbert space as
\begin{eqnarray*}
\widehat{H}_{NC}=B(w\p_{w}+1/2)+\widehat{V}(z,\bar{z}, w, \bar{w})
\end{eqnarray*}
in the NC case, and as
\begin{eqnarray*}
\widehat{H}_{E}=(B/m^*)\big(w\p_{w}+1/2\big)+
\widehat{V}(z,\bar{z}, w, \bar{w})
\end{eqnarray*}
in the ``exotic'' case, respectively.
Restricting our attention to the  subspaces $\widehat{z}=0$ and
$\p_{w}f=0$, respectively, we see
that   $f_{n}(w)=w^n$ (resp. $f_{n}(z)=z^n$) are simultaneous
eigenfunctions of both terms, yielding the total eigenvalues
$E_{n}=B(n+1/2)+\epsilon_{n}$ in the standard NC case,
and
$E_{n}=B/2m^*+\epsilon_{n}$ in the exotic case,
where $\epsilon_{n}$ is the eigenvalue of
$\widehat{V}(z,\bar{z},w,\bar{z})$, restricted to the respective
subspaces.

Note that $\epsilon_{n}$ is not yet known, since
$\widehat{V}(z,\bar{z},w,\bar{z})$ has not yet been calculated.
Letting $m^*$ go to zero, however, one or other of the
variables $z$ and $w$ become redundant, and $V$ takes the same form
$V(Z,\bar{Z})$.
The NC spectrum becomes hence
$
E_{n}=B(n+1/2)+\epsilon_{n}^{\rm crit},
$ cf. \cite{BNS}, while
the exotic spectrum is instead in the LLL,
$
E_{n}=(B/2m^*)+\epsilon_{n}^{\rm crit},
$
as suggested by Peierls. In the latter case
the first term diverges as $m^*\to0$, though,
and has to be removed by hand.
A similar behaviour has been observed by Dunne et al. \cite{DJT}
in the oscillator case $N=1$ who found that their $m\neq0$
energy eigenvalues diverge.

It is worth noting that the projection to the Hilbert
subspace (\ref{NCcondition}) can also be obtained by Bargmann-Fock
quantization of the $w$ plane, endowed with its canonical
symplectic structure and Hamiltonian,
\begin{equation}
     \begin{array}{ll}
	\Omega_{NC}^{\red}=(2i)^{-1}d\bar{w}\wedge dw,\hfill
	\\[8pt]
	H_{NC}^{\red}=\half Bw\bar{w}+V(w,\bar{w}).\hfill
	\end{array}
	\label{wNCredsympham}
\end{equation}
In the ``exotic'' case, the projected theory can be obtained
directly from a similar study
in the $z$-plane \cite{DJT, DH}.

\section{Classical aspects}

Studying the classical mechanics
of the two models provides us with further insight.

\subsection{The standard case}

The motion of a NC particle is governed by  Hamilton's equations,
$\dot{\xi}=\{\xi,H\}$, associated with the Poisson bracket
(\ref{BNSrel}), i.e.,
\begin{equation}
     \begin{array}{ll}
     \dot{x}_{i}
     =p_{i}-\theta\varepsilon_{ij}E_{j},
     \\[6pt]
     \dot{p}_{i}
     =B\varepsilon_{ij}p_{j}+E_{i},
     \label{NChameq}
     \end{array}
\end{equation}
$\vec{E}=-2U'(r^2)\vec{x}$
in the radial case $V=U(r^2)$.
According to the first equation, the velocity, $\dot{\vx}$, and the
momentum, $\vp$, are not in general the same or even parallel; in
the second equation the Lorentz force involves
$\vp$ and not $\dot{\vx}$.
  Eliminating $p_{i}$, we get
\begin{equation}
     \ddot{x}_{i}=
     B\varepsilon_{ij}\dot{x}_{j}+m^*E_{i}-\theta\varepsilon_{ij}
     \dot{E}_{j}
     \label{NCcleqmot}
\end{equation}
with
\begin{eqnarray*}
\dot{E}_{i}=-4U''(r^2)x_{k}\dot{x}_{k}x_{i}-2U'(r^2)\dot{x}_{i}
\end{eqnarray*}
for $V=U(r^2)$. Let us now put
$
\varpi_{i}=p_{i}-B\varepsilon_{ij}x_{j}.
$
The equations of motion, (\ref{NChameq}),  imply that
\begin{equation}
\dot{\varpi}_{i}=(1-B\theta)E_{i}\equiv m^*E_{i},
\label{varpivar}
\end{equation}
so that for $m^*=0$ $\varpi_{i}$
becomes, for {\it any} $E_{i}$, a {\it constant of the motion}.
  The dynamics can, therefore,
be consistently restricted to the two-dimensional surface
\begin{equation}
     \varpi_{i}\equiv
     p_{i}-B\varepsilon_{ij}x_{j}=c_{i}.
  \label{NCconstraint}
  \end{equation}
  The equations of motion retain their
form (\ref{NChameq}), and
our constraint (\ref{NCconstraint}) with $c_{i}=0$
reproduces that of Bellucci et al.,
as our $\varpi_{i}$ becomes their $\pi_{i}$ in (\ref{casimir}).
What we have found is the classical counterpart of
the irreducibility of the $x_{i}$ representation \cite{Sochi}.

Curiously, the ``new'' conserved quantity
$\varpi$ is related to the {\it translational invariance}:
in the absence of an $\vE$-field, the NC system in a constant
$B$-field would plainly be
invariant w. r. t. $\vx\to \vx+\vgamma$,
with associated conserved linear momenta $\varpi_{i}/m^*$.
Adding an arbitrary electric field
breaks this symmetry in general;
the conservation of $\varpi$ is, however,
restored when $B=\theta^{-1}$, since, in the critical case,
the electric field decouples, cf. (\ref{varpivar}).
This restauration of the translational symmetry is quite remarkable,
since it holds for {\it any} electric field.
Let us remind the Reader to the crucial r\^ole played by
the restauration
of [discrete] translational symmetry corresponding to a torus
geometry in deriving the quantization of the
Hall conductivity \cite{QHE}.

In what follows we shall also consider $c_{i}=0$
for simplicity, although this is not mandatory: another
  reduced theory could be obtained for each value of
the constants $c_{i}$.
The general one is obtained by a straightforward
modifications ; the complex coordinate $z$ in
(\ref{NCcomplcoord})  should become, e. g.,
$(Bm^*)^{-1/2}
\Big((-i\varpi_{1}+\varpi_{2})-(-ic_{1}+c_{2})\Big)
$,
etc.

In the radial case  the particle moves along circles~:
the electric field is radial, so (\ref{NChameq}) implies that
$
\dot{r}=2x_{k}\dot{x}_{k}=2x_{k}p_{k}=0
$
upon use of (\ref{NCconstraint}) with $c_{i}=0$.

Eliminating the position
$\vx$  using the constraint (\ref{NCconstraint}) allows us to
view the force as a function of $p_{i}$ alone,
$
E_{i}=E_{i}(-\varepsilon_{jk}p_{k}/B),
$
$E_{i}=({2}/{B})\varepsilon_{ij}p_{j}U'(\vp^2/B^2)$ in the
radial case. Therefore, the second equation
in (\ref{NChameq}) is actually a
first-order equation for $p_{i}$ ;
the first equation of in (\ref{NChameq})
is merely a consequence of the constraint
and of the second equation in (\ref{NChameq}).
This latter is actually Hamilton's equation associated with the
reduced symplectic structure/Poisson bracket and Hamiltonian
defined on $\vp$-space,
\begin{equation}
     \begin{array}{ll}
     \Omega_{NC}^{\red}=
     \displaystyle\frac{1}{B}dp_{1}\wedge dp_{2}
     \qquad\hbox{i.e.}\qquad
     \{f,g\}_{NC}^{\red}=B\left(\displaystyle\frac{\p f}{\p p_{1}}
     \displaystyle\frac{\p g}{\p p_{2}}
     -
     \displaystyle\frac{\p g}{\p p_{1}}\displaystyle
     \frac{\p f}{\p p_{2}}
     \right),\hfill
     \\[12pt]
     H_{NC}^{\red}=\half\vp^2+V\big(-\varepsilon_{ij}p_{j}/B\big).\hfill
\end{array}
\label{NCredsympham}
\end{equation}
Note  that the reduced Hamiltonian is simply the restriction of the
``original'' expression to the constraint surface (\ref{NCconstraint});
note also that (\ref{NCredsympham}) is consistent with the complex
expressions in (\ref{wNCredsympham}).
In the radial case the equation of motions is simply
\begin{equation}
     \dot{p}_{i}=\widetilde{B}\varepsilon_{ij}p_{j},
     \qquad
     \widetilde{B}=
     B+\frac{1}{B}2U'(\vp^2/B^2).
     \label{redpeq}
\end{equation}
Thus, the potential is transmuted into a (generally position-dependent)
effective magnetic field $\widetilde{B}$, as
observed before in the quadratic case \cite{NaPo}.
In fact, the force term in (\ref{NCcleqmot}) is switched off,
and the $\dot{E}_{i}$ merely contributes to the Lorentz force by
yielding an effective magnetic field $\widetilde{B}$.
It follows that $\vp$ rotates uniformly in the
``hodograph'' ( $\equiv \vp$ )-plane with uniform angular velocity
$\omega=\widetilde{B}$.
By (\ref{NCconstraint}) the position, $\vx$, performs the same type of
motion. In terms of the complex coordinate
(\ref{NCcomplcoord}),
the equations of motion associated with (\ref{wNCredsympham})
are solved by
$
w(t)=e^{i\widetilde{B}t}\zeta,
$
with [$\sqrt{B}$ times] $\zeta$ the initial $\vp$.

\goodbreak
\subsection{The exotic case}

Turning to the exotic case, we focus our attention to the
differences with the NC model.
The equations of motion,
\begin{equation}
     \begin{array}{rcl}\displaystyle
m^*\dot{x}_{i}
&=&
p_{i}-\displaystyle\theta\,\varepsilon_{ij}E_{j},
\\[8pt]
\displaystyle
\dot{p}_{i}
&=&
B\,\varepsilon_{ij}\dot{x}_{j}+E_{i},
\end{array}
\label{EEL}
\end{equation}
[cf. (\ref{NChameq})] can also be presented as
\begin{equation}
m^*\ddot{x}_i
=
\Big(B\varepsilon_{ij}\dot{x}_j+E_i\Big)
-\theta\Big(\varepsilon_{ij}
\dot{E}_{j}-\dot{B}\dot{x}_{i}\Big)
\label{modlorentz}
\end{equation}
($\dot{B}=\partial_\ell{}B\dot{x}_\ell+\partial_t{}B$),
which is rather different from (\ref{NCcleqmot}).
(The results  here do {\it not} require
$B$ to be constant \cite{DH}).
In the critical case $m^*=0$
the system becomes singular; Hamiltonian reduction \cite{FaJa}
performed in \cite{DH} requires the Hall constraint
\begin{equation}
     p_{i}=\varepsilon_{ij}\frac{E_{j}}{B}
     \label{Hallconstraint}
\end{equation}
analogous to but in general different from (\ref{NCconstraint}).
Then the $4D$
phase space  reduces  to a two-dimensional one with coordinate
$z$, consistent with (\ref{Ecomplcoord}).
The classical phase space is hence the complex plane with
canonical symplectic structure $(2i)^{-1}d\bar{z}\wedge dz$,
and the reduced Hamiltonian is $H_{E}^{\red}=V=V(z, \bar{z})$ alone:
this is the classical counterpart of the Peierls substitution.
The second-order equations (\ref{modlorentz})
reduce to a first-order ones, namely to the {\it Hall law},
\begin{equation}
     \dot{Q}_{i}=\varepsilon_{ij}\frac{E_{j}}{B}.
\label{Hallaw}
\end{equation}
These latters can be obtained
from the reduced symplectic structure and Hamiltonian
\begin{equation}
     \begin{array}{ll}
     \Omega_{E}^{\red}=BdQ_{1}\wedge dQ_{2},
     \hfill\\[8pt]
     H_{E}^{\red}=V(Q_{1}, Q_{2})
     \end{array}
     \label{Eredsympham}
\end{equation}
where the $Q_{i}=x_{i}-E_{i}/B^2$ are suitable coordinates \cite{DH}.

Hence, consistently also with the conservation of the reduced energy
$H_{E}^{\red}=V$, the motions follow equipotentials.
For a radial potential $V=U(r^2)$ in particular, the trajectories are
again circles,
with (radius-dependent) uniform angular velocity $\omega=2U'(r_{0}^2)/B$
({\it different} from that in the NC case). In complex notations,
$z(t)=e^{-i\omega t}\zeta$, where
$
\omega=N\big(\zeta\bar{\zeta}/2\big)^{N-1}/{B}.
$

\subsection{Comparison of the models}

The difference between the models originates in the way the particle
is {\it coupled to the gauge field}.
In the standard NC case \cite{BNS, NaPo}, the rule is to {\it replace the
  commutation relation  of the momenta},
  $\{p_{1}, p_{2}\}=0$, by
  the last one in (\ref{BNSrel}),  viz. $\{p_{1}, p_{2}\}=B$.
  Note that this only works for a
  constant $B$, otherwise the posited Poisson bracket
  will not satisfy the Jacobi identity~: e.g.,
  $\big\{x_{i},\{p_{1},p_{2}\}\big\}_{\rm
cyclic}=\theta\varepsilon_{ij}\p_{j}B$.
  This is an unpleasant restriction, since the requirement
  of having a strictly constant $B$ is rather unphysical.

  The  recipe followed in the exotic case is instead that of
  Souriau \cite{SSD}, who first unifies both the symplectic structure
  and the Hamiltonian into a single two-form, viz.
  $\sigma=\Omega-dH\wedge dt$. Then his rule says that
  the minimally coupled
  two-form should be obtained by adding the electromagnetic tensor
  $F$ to the free two-form $\sigma_{0}$.
  In this framework, the Jacobi identity holds
for any gauge field: it comes from that $\sigma_{E}$ is a closed
$2$-from, $d\sigma_{E}=0$,  which follows in turn
from the homogeneous Maxwell equation $dF=0$.

The two rules are only equivalent for $\theta=0$ or for $B=0$,
as it can be readily seen remembering that
the Poisson bracket involves the {\it inverse}
of the  symplectic matrix $\Omega$,
\begin{equation}
     \{f,g\}=P^{\alpha\beta}\p_{\alpha}f\p_{\beta}g,
     \qquad
     P^{\alpha\beta}=\big(\Omega^{-1}\big)_{\alpha\beta}.
     \label{PBdef}
\end{equation}

  Explicitely, Souriau's rule yields
  \begin{equation}
      \sigma_{E}=d\vp\wedge d\vx+\theta dp_{1}\wedge dp_{2} +
     B dx_{1}\wedge dx_{2}-(\vp\cdot d\vp+dV)\wedge dt,
     \label{Esigma}
\end{equation}
whereas the Poisson bracket posited in the NC approach
  corresponds to the manifestly different two-form
  \begin{equation}
      \sigma_{NC}=
      \frac{1}{m^*}
      \Big[d\vp\wedge d\vx+\theta dp_{1}\wedge dp_{2} +
     B dx_{1}\wedge dx_{2}\Big]
     -(\vp\cdot d\vp+dV)\wedge dt.
     \label{NCsigma}
\end{equation}
\goodbreak

\subsection{Variational aspects}

A new light  is shed on the two models by
studying the variational aspects.
The classical action can in fact be written as the integral of
Cartan's $1$-form, $\int Ldt=\int\Theta$ \cite{SSD}; then the
associated Euler-Lagrange equations say that the motions curves
are tangent to the kernel of the $2$-form $\sigma=d\Theta$ \cite{SSD}.
For
$\sigma=\Omega-dH\wedge dt$, this means
\begin{equation}
\Omega_{\alpha\beta}\dot{\xi}_{\beta}=\frac{\p H}{\p\xi_{\alpha}}
\label{ELeq}
\end{equation}
($\xi_{\alpha}=(p_{i}, x_{j})$). When the matrix
$\Omega_{\alpha\beta}$ is regular,
(\ref{ELeq}) can be inverted, and we get Hamilton's equations, written
in terms of the Poisson bracket (\ref{PBdef}) as $\dot{\xi}_{\alpha}=
\big\{\xi_{\alpha}, H\big\}$.
When  $\Omega_{\alpha\beta}$ is singular, however, one can only
derive a Poisson bracket-formulation after Hamiltonian \cite{FaJa}
(alias symplectic \cite{SSD}) reduction.
Conversely, when a Poisson bracket
and a Hamiltonian
are  posited, one can only reconstruct a Lagrangian
provided the matrix $P^{\alpha\beta}$ which defines the Poisson bracket
is regular.

In the NC case the posited Poisson structure (\ref{BNSrel})
only leads to the $2$-form (\ref{NCsigma}) and hence to a
variational formulation {\it off} the critical case. In fact
\begin{equation}
     \Theta_{NC}=\frac{1}{m^*}
     \left[(p_{i}-A_{i})dx^i+
     \half\theta\varepsilon_{ij}p_{i}dp_{j}\right]
     -(\half{\vp}^{2}+V)dt
\label{NCcartan}
\end{equation}
works (contrary to claims \cite{ACAT}) when $m^*\neq0$.
It blows up, however, when $m^*\to0$ --- although
   Hamilton's equations behave regularly. This latter can hence
   only by derived from a variational principle
after reduction. The Hamiltonian structure
(\ref{NCredsympham}) corresponds indeed to the first-order
Cartan $1$-form [Lagrangian]
\begin{equation}
     \Theta_{NC}^{red}=\frac{1}{2B}\varepsilon_{ij}p_{i} dp_{j}
     -\Big(\half\vp^2+V(-\varepsilon_{ij}p_{j}/B)\Big)dt,
     \label{NCredcartan}
\end{equation}
cf. \cite{ACAT}.
The exotic Cartan form is instead
\begin{equation}
     \Theta_{E}=(p_{i}-A_{i})dx^i+
     \half\theta\varepsilon_{ij}p_{i}dp_{j}
     -(\half{\vp}^{2}+V)dt,
\label{Ecartan}
\end{equation}
whose exterior derivative, $d\Theta_{E}=\sigma_{E}$,
becomes singular at the critical point $m^*=0$.
  Then there is no associated Poisson bracket
structure, and a Hamiltonian formulation is only possible after
``Faddeev-Jackiw'' reduction \cite{FaJa, DH}. These latter
are consistent with the
variational $1$-form
\begin{equation}
     \Theta_{E}^{red}=\frac{B}{2}\varepsilon_{ij}Q_{i} dQ_{j}
     -V(Q_{1}, Q_{2})dt,
     \label{Eredcartan}
\end{equation}
used before by Dunne at al. in their $m\to0$
derivation of the Peierls rule \cite{DJT}.

\section{Vortex dynamics in superfluid helium}

The  interest of the Peierls-type
model is underlined by that it also describes the
effective dynamics of point-like flux lines in a thin film of
superfluid ${}^4$He \cite{HMC, LeMy}.
For the sake of simplicity, we restrict ourselves to  two
vortices of identical vorticity.
The center-of-vorticity coordinates are constants of the motion.
Introducing the relative coordinates
$x=x_{1}-x_{2}$ and $y=y_{1}-y_{2}$ and the respectively,
the vortex equations become \cite{HMC}
\begin{eqnarray}
     (\rho\delta\kappa)\frac{dx}{dt}=\frac{\p H}{\p y},
     \qquad
     (\rho\delta\kappa)\frac{dy}{dt}=-\frac{\p H}{\p x},
      \label{effdyn}
\end{eqnarray}
where $\rho$ and $\delta$ are the density and the thickness of the
film, respectively;
$\kappa$, the vorticity,  is quantized in units
  $h/m$, where $h$ is Planck's
constant and $m$ is the mass of the helium. (In what follows we choose
units where $\hbar=1$). The effective Hamiltonian here has the form
\begin{equation}
H=-(\rho\delta\kappa^2/4\pi)\ln\big[x^2+y^2\big].
\label{effham}
\end{equation}

The system has fractional angular momentum and
  obeys fractional statistics \cite{HMC, LeMy}.

A key observation for our purposes \cite{LeMy} is that
(\ref{effdyn}) is in fact a Hamiltonian system, $\dot{\xi}=\{\xi,H\}$
($\xi=(x,y)$),
consistent with
the Poisson bracket associated with the symplectic structure
  $\theta^{-1}dx\wedge dy$, $\theta^{-1}=\rho\delta\kappa$
of the plane $(x,y)$, cf.
(\ref{Eredsympham}) with $B=\theta^{-1}$.
Vortex dynamics in superfluid helium provides us hence
with yet another physical instance of non-commuting coordinates.
Let us emphasize that the symplectic plane should be viewed as
the classical {\it phase space} for the vortex motion.

\goodbreak
Another peculiarity is the absence of a mass term in the hamiltonian
(\ref{effham}).
The analogy with the motion of {\it massless} particles in a
magnetic field, noticed \cite{HMC, LeMy} before,
can be further amplified:  (\ref{effham}) is indeed the ``reduced
Hamiltonian''  $H^{\rm red}_{E}\propto\ln r^2$ in (\ref{Eredsympham}).

The identity of the underlying mathematical structures
allows us to transfer the analysis in presented above to the
superfluid case.
For any radial Hamiltonian $H=U(r^2)$ in particular,
the equipotentials are circles, so the relative motion of the
vortices is a rotation
with (separation-dependent) uniform angular velocity
$\omega=2U'(r_{0}^2)/\rho\delta\kappa$.
For the effective vortex Hamiltonian (\ref{effham}) in particular,
the angular velocity is inversely proportional to the square of the radius,
$\omega=-(\kappa/2\pi)r^{-2}$. (This is also consistent with the
conservation of vorticity).

As explained above, quantization is conveniently carried out in the
Bargmann-Fock
framework.
The spectrum can be found by quantizing
  $H=-(\rho\delta\kappa^2/4\pi)\ln r^2$. Leinaas and Myrheim claim
that it is simply
\begin{equation}
\epsilon_{n}=-\frac{\rho\delta\kappa^2}{4\pi}\ln(8\theta\, j_{n}),
\label{LeMyspectrum}
\end{equation}
where the $j_{n}$ are the eigenvalues of the
  conserved angular momentum
$
J=\half\rho\delta\kappa r^2=\half z\bar{z}.
$
This ignores the ordering problem, and
agrees with the first quantization scheme discussed in
Sect. \ref{quantization}.
Although we have not been able to
calculate it explicitely, the answer provided the other schemes
will definitely be different from
(\ref{LeMyspectrum}).

Leinaas and Myrheim derive the spectrum of the angular momentum
  from the representation theory. They observe in fact
that $J=A/2$, where the
\begin{eqnarray}
     A=(z\bar{z}),
     \qquad
     B=\frac{1}{2}(z^2+\bar{z}^2)
     \qquad
     C=\frac{1}{2i}(z^2-\bar{z}^2)
     \label{dynalgebra}
\end{eqnarray}
span an sl$(2,\IR)\simeq{\rm o}(2,1)\simeq$\,sp$(1)$ algebra,
whose representation
theory yields  the fractional eigenvalues
$j_{n}=(\alpha_{0}+n)$ of the angular momentum \cite{LeMy}.

The arisal of this algebra is easy to
understand: the operators in (\ref{dynalgebra}) are the generators
of the symplectic group sp$(1)$ in the plane.
We argue, however, that this only works for the angular
momentum, but not for a general Hamiltonian, since
$H=U(r^2)$ does not belong to sp$(1)$,
except in the trivial case $U(r^2)\propto r^2$, and going from
$r^2$ to $U(r^2)$ is precisely the problem discussed in
Sect. \ref{quantization}.
Its spectrum can not
be derived therefore from the representation theory of sp$(1)$ in general.

\section{Infinite symmetry}

The point is that sp$(1)$ belongs to a much
larger --- actually infinite-dimensional --- symmetry algebra,
found before for the edge currents
in the Quantum Hall context  \cite{Winfty}.
The reduced system (\ref{Eredsympham})
is, namely, symmetric w. r. t.
  $w_{\infty}$, the algebra of all area-preserving diffeomorphisms.
   This is the best explained in terms of Souriau's
``{\it espace des mouvements}'' [= space of motions]).
Here we only present a brief outline;
for details the Reader is referred to
Souriau's book \cite{SSD}.
The space of motions we denote here by ${\cal M}$
consists of entire motion curves of the system. The classical
dynamics is encoded into the symplectic form of ${\cal M}$ we denote
by $\sigma$. This latter incorporates both the phase space
symplectic structure and the Hamiltonian~: in fact, $\sigma$ is the
projection along the motion curves of the closed two-form $\sigma=
d\Theta=\Omega-dH\wedge dt$ [justifying our abuse of notation].

It is then obvious that {\it any} function $f(\zeta)$ on ${\cal M}$ is
a constant of the motion, since it only depends on the motion $\zeta$.
Integrating the equations of motions, $f(\zeta)$ can be expressed using
the phase space variable $\xi$ and time, $t$,
$f(\xi, t)=f(\zeta)$;
such a function can look rather complicated, though,
owing to the complicated relation between $\zeta$ and $\xi, t$.
Its conservation means
\begin{equation}
     \p_{t}f=\{f,H\},
\label{conservation}
\end{equation}
where $\{\cdot,\cdot\}$ is the Poisson bracket associated
with the phase-space symplectic form $\Omega$.

To any such function $f(\zeta)$ corresponds,through
$
-\partial_{\mu}f=\sigma_{\mu\nu}Z^\nu,
$
a (``Hamiltonian'')  vectorfield $Z^\mu$,
which generates in turn, at least locally, a $1$-parameter
group of diffeomorphisms of ${\cal M}$ which leaves
$\sigma$ invariant.
All such diffeomorphisms
form an infinite dimensional group, namely the group of
symplectomorphisms  of ${\cal M}$. Any symplectic transformation is a
symmetry of the system : it merely permutes the motions. This group is
generated by all expressions of the form $\bar{z}^nz^m$, generalizing
the quadratic subalgebra (\ref{dynalgebra}).

For any radial potential $U(r^2)$  the initial position
$\zeta=e^{i\omega t}z(t)$
(where $\omega$ is the angular velocity
calculated above)
is a good coordinate on the space of motions.
We find furthermore that
\begin{equation}
     \sigma=\frac{d\bar{z}\wedge dz}{2i}-dH\wedge dt
     =\frac{d\bar{\zeta}\wedge d\zeta}{2i},
     \label{Zsymp}
\end{equation}
proving that the space of  motions is, for {\it any} radial potential,
the symplectic $\zeta$-plane.
But in two dimensions, the $\Omega$-preserving transformations
are the same as the {\it area preserving
diffeomorphisms}. The Lie algebra structure is defined by the
Poisson bracket associated with (\ref{Zsymp}).

The Hamiltonian, the generator of time translations,
is a constant of the motion, and is hence defined
on ${\cal M}$. It
belongs therefore to our algebra $w_{\infty}$. The spectrum of the
corresponding
quantum operator should  result therefore also from the representation
theory of the quantum-deformed version of $w_{\infty}$.

\section{Conclusion}

In this paper we studied the Landau problem in
two  non-commutative frameworks. The models
are equivalent in the free case \cite{DH},
but lead to different (albeit similar)  conclusions when interactions
are introduced.
The difference comes from the way the gauge coupling is defined.
In the standard NC approach \cite{BNS, NaPo, GAMB},
the Poisson bracket ---  a {\it contravariant} structure --- is modified;
in the Peierls-type one \cite{DJT, DuJa, DH}, the coupling is achieved
using Souriau's {\it covariant} two-form $\sigma$.
The two ``minimal coupling'' rules are hence the {\it duals} of each other.
The first one yields a complete Landau-type specturm, and the  ``exotic''
one only yields LLL states.
The latter applies to
the Fractional Quantum Hall Effect. The standard NC framework may be
relevant instead for the Integer Effect \cite{QHE, MorPo}
\goodbreak

  Remarkably, the ``NC'' Poisson bracket
(\ref{BNSrel}) and the ``exotic'' two-form (\ref{Esymp})
[or (\ref{Esigma})]
become both singular for the same critical value $m^*=0$,
necessitating reduction from $4D$ to $2D$ phase space.
In the NC case the reduced manifold corresponds to fixing the value of
the conserved linear momentum $\varpi$, and is parametrized by the
canonical momentum, $\vp$. The reduced dynamics is given by
(\ref{NCredsympham}). In the exotic case
we get a ``coordinate'' picture, with dynamics (\ref{Eredsympham}).
The motion obeys the Hall law.
The trajectories in the two theories
are similar; the difference arises owing to the
extra kinetic term in the (reduced) Hamiltonian $H_{NC}^{\red}$.

A singular Poisson structure with related variational
problems has been exhibited in hydrodynamics
\cite{JACKREV}, and in the study of
quantum Hall fluids \cite{QHF}. The models discussed in this paper
provide further examples.

The fact that the same simple Peierls-type model also describes superfluid
vortex dynamics  underlines the physical importance of a classical
study. Our results provide further evidence for the
  intimate relation between the fractional Quantum Hall Effect
and superfluidity \cite{LeMy}.
Let us remind the Reader that $w_{\infty}$ is also the symmetry of
  incompressible fluids \cite{Arnold} in general
  and of Quantum Hall fluids in particular \cite{Winfty, QHF}.

  It is also worth mentionning that
  this same infinite-dimensional symmetry is the starting point in another,
  related approach \cite{Stichel}.

  \kikezd{\bf Acknowledgement.}
I am indebted to Professors
J. Balog, G. Dunne, J. Gamboa, R. Jackiw, J.-M. Leinaas, L. Martina,
G. Tuynman, and E. Varoquaux for correspondence,
  and to M. Niedermayer and F. Schaposnik for discussions.
  Special thanks are due to C. Duval for his
  constant advice, help and interest.


\end{document}